\documentclass{IEEEtran}  % Comment this line out[12pt,draftcls,onecolumn]
\usepackage{graphics} % for pdf, bitmapped graphics files
\usepackage{epsfig} % for postscript graphics files
\usepackage{mathptmx} % assumes new font selection scheme installed
\usepackage{times} % assumes new font selection scheme installed
\usepackage{amsmath} % assumes amsmath package installed
\usepackage{amssymb}  % assumes amsmath package installed
\usepackage{amsfonts}
\usepackage{diagbox}
\usepackage{amsthm}
\newtheorem{theorem}{Theorem}[section]
\newtheorem{corollary}{Corollary}
\newtheorem{lemma}[theorem]{Lemma}
\newtheorem{remark}{Remark}
\newtheorem{definition}{Definition}
\usepackage{enumerate} 
\usepackage[usenames, dvipsnames]{color}
\usepackage{booktabs}
\usepackage{rotating}
\usepackage{setspace}

\usepackage[margins]{trackchanges}

\title{\LARGE \bf
Input-Output Stability of Barrier-Based Model Predictive Control}

\author{Panagiotis Petsagkourakis,~\IEEEmembership{Student Member,~IEEE}, William P. Heath,~\IEEEmembership{Member,~IEEE}, Joaquin Carrasco,~\IEEEmembership{Member,~IEEE} and Constantinos Theodoropoulos% <-this % stops a space
}

\addeditor{PP}
\addeditor{WPH}

\begin{document}
\maketitle

\begin{abstract}
Conditions for input-output stability of barrier-based model predictive control of linear systems with linear and convex nonlinear (hard or soft) constraints are established through the construction of integral quadratic constraints (IQCs). 
%are constructed for the nonlinearity associated with barrier-based MPC with convex input constraints that can be hard or soft. 
The IQCs can be used to establish sufficient conditions for global closed-loop stability. In particular conditions for robust stability can be obtained in the presence of unstructured model uncertainty. IQCs with both static and dynamic multipliers are developed and appropriate convex searches for the multipliers are presented. %for the case of convex input constraints with static and dynamic multipliers.
The effectiveness of the robust stability analysis is demonstrated with an illustrative numerical example.%, employing an effective convex search. 
\end{abstract}

%%%%%%%%%%%%%%%%%%%%%%%%%%%%%%%%%%%%%%%%%%%%%%%%%%%%%%%%%%%%%%%%%%

\section{Introduction}
Model predictive control (MPC) has been widely used to compute a sequence of control inputs online by optimizing an objective function with suitable constraints in a receding horizon manner. Nevertheless, it remains hard to guarantee robustness without introducing prohibitive complexity~\cite{Mayne2000}. 
%In this paper, we use a tool for classic robust stability, namely input-to-output stability. 
In this paper we obtain sufficient conditions for global robust stability for barrier-based MPC with linear or convex nonlinear input constraints where the plant is open-loop stable through the use of integral quadratic constraints (IQCs). IQCs, introduced in~\cite{Yakubovich1967}, provide a unified framework for analyzing the robustness of Lur'e type systems. 

Classic robust control theory deals with stability and robustness under unstructured uncertainty, where IQCs are used to represent nonlinear and uncertain components ~\cite{Megretski1997}.  Recent  studies  include robust stability analysis and perfomance ~\cite{Veenman2016}, for LPV ~\cite{Pfifer2015} and for distributed parameter systems ~\cite{Cantoni2009}. Jonsson and Rantzer~\cite{Jonsson2000} illustrate how IQCs can be
used to analyze robust stability of anti-windup control systems with both unstructured uncertainty and nonlinearities. 
Passivity, dissipativity and IQCs have been explored to analyze the input-output stability of MPC.  
Robust stability analysis of MPC has been performed by utilizing IQCs in ~\cite{Heath2006}. Also, a robust output  MPC design based on dissipativity for unstructured uncertainties has been proposed proposed ~\cite{Lovaas2008}. More recently IQCs in conjunction with dissipativity have been used for multi-model MPC~\cite{Petsagkourakis2018} and later for LPV based MPC~\cite{Cisneros2018}. To overcome the extensive conservatism for the analysis of MPC, the existence of Zames-Falb (ZF) multipliers was shown ~\cite{Heath2007a}, originally proposed in~\cite{Zames1967}, for time-invariant linear constraints. 
%and recently more general multipliers have been found for a more specific class of constraints~\cite{Heath2010}. 
%The use of ZF multipliers was proposed in~\cite{Zames1967} and they can reduce the conservatism of the stability analysis significantly. 
In \cite{Safonov2000} necessary conditions have been provided for the existence of such multipliers for the case of MIMO nonlinearities. Additionally, repeated nonlinearities have been treated in~\cite{Damato2001} where a special class of multipliers has been proposed. In ~\cite{Jonsson2000} a unified framework for multiplier search by LMI optimization was formulated. 
Recently,  a comprehensive analysis for the case of slope-restricted nonlinearities in discrete time was proposed ~\cite{Fetzer2017b, Carrasco2018_search, Wang2014}.

Barrier-based MPC was proposed in~\cite{Wills2004a} where stability of state feedback MPC can be established using gradient recentering; its implementation on an edible oil refining plant was reported in~\cite{WILLS2005183}. Fast implementations of MPC using barriers are reported in~\cite{Wang2010a}. Recently, in a series of papers~\cite{Feller2015,feller2017relaxed, Feller2017}  barrier-based MPC is developed to establish stability results for numerically efficient and practical MPC implementations. 
Innovations of these works include  weight-recentered barriers, analysis of relaxed barriers for soft constraints and stability results for {\it anytime algorithms} (where the number of Newton steps in the associated optimization algorithm may be small).
In this paper, we  use recentered barriers  for  hard  constraints  and a relaxed  recentered  barrier  for  soft  constraints.  This  complements  the  work in  ~\cite{Feller2017,Feller2016} which considers  stability  of  state  feedback  MPC for soft input and state constraints. Here, stability analysis allows hard input constraints which may be time variant or nonlinear. 
In the case of time invariant constraints, ZF multipliers can be used. Hence, the search for such multipliers becomes crucial.
% 
%
%
%In this paper, stability of output feedback MPC with convex input constraints is considered. 
%It is shown that as the design parameter $\mu$ increases the predicted ({\it and actual}) stability region. 
%The stability analysis is performed with respect to multipliers, as a result the search of such MIMO multipliers becomes crucial. 
%where the slopes are given by diagonal matrices.  
%showing that most stability tests in the literature are related  
%
%For more details regarding the ZF multipliers and their application, there are two comprehensive tutorials in~\cite{Carrasco2015} and~\cite{Fetzer2017b}.  
%In this work we extend the methodologies in ~\cite{Fetzer2017b, Carrasco2018_search} by proposing a general convex search for multitpliers, when the slopes are %symmetric matrices instead of diagonal
%~\cite{Damato2001}. 
%
In this paper we construct static and dynamic multipliers, exploiting convex searches, for barrier-based MPC, which are then used for input-to-output stability and robust analysis, illustrating the advantages that barrier MPC can provide. 

In section~\ref{Not} the basic notations used in the paper are provided. In section~\ref{Prob} the formulation of barrier MPC is discussed and in section~\ref{prelim}, some basic results are presented. In section~\ref{properties} and~\ref{multipl} our main results are presented. The properties of barrier MPC are investigated, and the existence of static/ dynamic multipliers is shown.  In section~\ref{search}, a convex search methodology for efficiently computing multipliers is presented. In section ~\ref{examples}  an illustrative numerical example is shown, while conclusions are  given in ~\ref{conc}. 
%the conclusion of this work are given.
%\begin{center}\label{la}
%\begin{tabular}{ c c c c }
%\hline
%\backslashbox{stability}{constraints} & Barrier & gradient recentred barrier & %relaxed gradient recentred barrier\\ 
% \hline
%{circle criterion} & $\times$ & $\checkmark$ & $\checkmark$ (for only state %constr)\\  
%ZF & $\times$ & $\checkmark$ & $?$   \\
% \hline 
%\end{tabular}
%\end{center}
%%%%%%%%%%%%%%%%%%%%%%%%%%%%%%%%%%%%%%%%%%%%%%%%%%%%%%%%%%%%%%
\section{Notation}~\label{Not}
Let $l^{m}$ be the space
of  all  real-valued  sequences. $\mathbf{RH}_{\infty}$ is the set of rational matrix transfer function matrices without poles outside the unit circle. Let $x_k\in {\rm I\!R}^{n_x}$ be the value of $x\in l^{n_x}$ at sample $k$. Let $A^{*}$ be the complex conjugate transpose of the matrix $A$ and let $G^{*}$ be the $l_{2}$-adjoint operator of $G$. $\left\langle f,g \right\rangle$  is the inner product of real-valued sequences $f$ and $g$, defined as $\sum_{k=-\infty}^{\infty}f_{k}^{\intercal}g_{k}=\frac{1}{\pi}\int_{-\pi}^{\pi} \hat{f}(e^{j\omega})\hat{g}(e^{j\omega})d\omega$, $\hat{f}$ being the Fourier transform of $f$. $\sqrt{\left\langle f,f \right\rangle}$ is the $l_{2}$ norm $\Vert f\Vert _{2}$. The discrete convolution at time $i$ is notated as $(f*g)_i = \sum_{k={-\infty}}^{\infty}f_kg_{i-k}$. The size of signal $x$ is $n_x$, while $I$ denotes the identity matrix. 
%and $\mbox{diag}(A,B)$ denotes the block diagonal matrix $\begin{bmatrix}A& {} \\ {} & B\end{bmatrix}$. 
%of appropriate size. 
For a matrix $A\in {\rm I\!R}^{m\times n}$ with rank $r$ we define $A^c\in {\rm I\!R}^{n-r\times n}$ such that $A^cA^T=0$, $A^cA^{cT}=I$, and $\bar{A} \in {\rm I\!R}^{r\times n}$ such that
\begin{equation} \label{eqnot}
\begin{split}
\bar{A} = \begin{cases}
({A^{c}})^c~~~, \textrm{when} ~ r<n\\
I~~~, \textrm{when} ~ r=n
\end{cases}
\end{split}
\end{equation}
Hence, $\bar{A}\bar{A}^T=I$ and the rows of $\bar{A}$ form an orthonormal basis of the space spanned by the rows of $A$. Furthermore, the set of all sub-gradients of a function $f$ at $x$ is called the $\it subdifferential$ of $f$ at $x$ and it is denoted by $\partial f= \partial f(x)$.
% Furthermore $\mathcal{U}^o$ is defined as the open interior of $\mathcal{U}$.
%%%%%%%%%%%%%%%%%%%%%%%%%%%%%%%%%%%%%%%%%%%%%%%%%%%%%%%%%%%%%%%%%%%%%%555
%%%%%%%%%%%%%%%%%%%%%%%%%%%%%%%%%%%%%%%%%%%%%%%%%%%%%%%%%%%%%%
\section{Formulation of Barrier MPC}~\label{Prob}
%\begin{figure}
%\centering
%\includegraphics[scale=0.3]{Picture1s.png}
%\caption{Linear system under uncertainty}
%\label{fig:linear}
%\end{figure}
The controller is designed using the nominal LTI model (without any information for the disturbances). The nominal system is modelled as 
\begin{equation}\label{model}
\begin{split}
&x_{k+1} = Ax_k+B_u u_k\\
&y_k = C x_k\\
\end{split}
\end{equation}
where $x_k\in {\rm I\!R}^{n_x}$ is the vector of states, $u_k\in {\rm I\!R}^{n_u}$ the vector of manipulated variables, and $y_k\in {\rm I\!R}^{n_y}$
 the vector of measured output variables.
%and $w_k\in {\rm I\!R}^{n_w}$, and $v_k\in {\rm I\!R}^{n_v}$ are the vector of additional inputs and outputs, of the causal operator $\Delta$. This system is depicted in Figure~\ref{fig:linear}. Here, we 
Additionally, we assume that $A\in {\rm I\!R}^{n_x\times n_x}$ is Schur stable. %and the interconnection is well-posed. 
%%%-------------------------------------------------------------
%%%
%%%
%%%
The formulation of barrier MPC is deduced from the nominal constrained problem, hence nominal MPC is presented first.  
The control action for the nominal problem is given as:
\begin{subequations}\label{conv}
\begin{equation}\label{obj_c}
    U_k =\arg\min_{\tilde{u}_k} \frac{1}{2}[\sum_{i=1}^{N}\hat{x}_{k+i|k}^T Q \hat{x}_{k+i|k} +\sum_{i=0}^{N-1}\hat{u}^T_{k+i|k} R\hat{u}_{k+i|k}]
\end{equation}
\begin{equation}\label{equal_conv}
\begin{split}
       & s.t.~\hat{x}_{k|k} = x_k\\
     &\hat{x}_{k+i|k} = A\hat{x}_{k+i-1|k}+B_u \hat{u}_{k+i-1|k}\\
&\hat{y}_{k+i-1|k} = C \hat{x}_{k+i-1|k}\\   
\end{split}
\end{equation}
\begin{equation}\label{ineq_conv}
\tilde{u}_k=\begin{bmatrix}\hat{u}_{k|k}^T & \dots&\hat{u}_{k+N-1|k}^T\end{bmatrix}^T\in \mathcal{U}
\end{equation}
\end{subequations}
with $N$ the prediction horizon. The non-empty compact convex set $\mathcal{U}$ represents the input constraints, $F_i: {\rm I\!R}^{n\times n_u} \rightarrow {\rm I\!R}$:
\begin{equation}
\mathcal{U}:\left \{F_i(\tilde{u})\leq W_i |i\in\left\{1,\dots,n\right\}\right \}
\end{equation}
where $F$ is a convex function. For the specific case of linear inequalities, the constraints can be written as $F(\tilde{u})=L~\tilde{u}\leq W${, assuming $F_i(0) = 0$ and $W_i \geq 0 $}. If we add the equality constraints (3b) to the objective function (3a) we obtain:

\begin{equation}\label{conv1}
\begin{split}
&U_k =\arg\min_{\tilde{u}} \frac{1}{2}\tilde{u}^T H \tilde{u} - \theta_k^T \tilde{u}\\
&\tilde{u}\in \mathcal{U}
\end{split}
\end{equation}
where $H, R, Q$ can be defined as in~\cite{maciejowski2002predictive} and $\theta$ is a linear function of states ($\theta_k = -Sx_k$).
The constrained problem is then transformed to an unconstrained problem according to \cite{Feller2015,Wills2004a} using barrier functions and/or penalty (relaxed barrier). 
%The function $B:\mathcal{U}^o\rightarrow {\rm I\!R}$ is a ${\vartheta}$-self-concordant barrier~\cite{Nesterov1994} defined on an open non-empty convex subset $\mathcal{U}^o$ and the recentered gradient  is defined as follows:
\begin{figure}[t]
\centering
\includegraphics[scale=0.35]{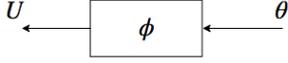}
\caption{Input-to-output map of Barrier MPC}
\label{fig:phi}
\end{figure}
\begin{definition}\label{def1}
Let $\mathbf{U}$ be an open (strictly) convex
set which contains the origin, defined as $\mathbf{U} := \cap \mathbf{U}_i$, where $\mathbf{U}_i$ is $\left\{\tilde{u}:  F_i(\tilde{u})\leq W_i\right\}$ for the hard constraints or ${\rm I\!R}^{n_U}$ for the soft constraints.   Then $\mathcal{B}$ is the set of all twice continuous differentiable ${\vartheta}$-self-concordant  (strictly) convex barrier functions over $\mathbf{U}$, ${B}:\mathbf{U}\rightarrow {\rm I\!R}$  with $B(0) = 0$ and $\nabla B(0) = 0$. 
\end{definition}
In the literature there are two popular barrier functions that are utilized. The {\it gradient recentered log-barrier}~\cite{Wills2004a} 
\begin{equation}\label{grad}
    B(U) = \sum_i B_i = \sum_i \left( -\ln(W_i-F_i(U)) + \ln(W_i)-\dfrac{\nabla F_i(U)^T}{W_i-F_i(0)}U\right) 
\end{equation}
%%%%%%%%%%%%%
and the {\it weighted  recentered log-barrier}~\cite{Feller2015}.
\begin{equation}\label{weig}
    \begin{split}
        &B(U) = \sum_i B_i = \sum_i (1+w_i)\left( -\ln(W_i-F_i(U)) + \ln(W_i)\right) 
    \end{split}
\end{equation}
with $w_i>0$. In addition, Feller and Ebenbauer~\cite{Feller2015} have proposed a relaxed barrier function to be employed, substituting the natural logarithm in (\ref{grad}) \& (\ref{weig}) with a quadratic function $\beta_i$ when $W_i-F_i(U)\leq \delta_i$ with $\delta_i>0$. The quadratic function is defined so that the properties from Definition~\ref{def1} are maintained but the domain of the function $B_i$ is ${\rm I\!R}^{n_U}$. 
%
%Define a function $B:\mathbf{Q}\subset {\rm I\!R}^{n_U}$ as
%\begin{equation}\label{prog}
%B(U)=\bar{B}(U)-\bar{B}(0)-\nabla \bar{B}(0)^T~U
%\end{equation}
%Then $B$ is called gradient recentred ${\vartheta}$-self-concordant barrier
%function over $\mathbf{Q}$ (about the origin).
The use of a barrier function allows the elimination of the inequality constraints. 
%\begin{equation}\label{conv2}
%\begin{split}
%&U =\arg\min_{{u}_k} \frac{1}{2}[\sum_{i=1}^{N}\hat{y}_{k+i|k}^T q \hat{y}_{k+i|k} +r\sum_{i=0}^{N-1}\hat{u}^T_{k+i|k} \hat{u}_{k+i|k}]+\mu B(\hat{u}_k)\\
%&s.t. \\
%&\hat{x}_{k+i|k} = A\hat{x}_{k+i-1|k}+B \hat{u}_{k+i-1|k}\\
%&\hat{y}_{k+i-1|k} = C \hat{x}_{k+i-1|k}\\
%&\hat{u}_k=\begin{bmatrix}\hat{u}_{k|k}\\ \vdots\\\hat{u}_{k+N-1|k}\end{bmatrix}
%\end{split}
%\end{equation}
\begin{equation}\label{cycmin}
U_k=\phi (\theta_k) =\arg\min_{\tilde{u}} \dfrac{1}{2}\tilde{u}^TH\tilde{u}-\theta_k^T~\tilde{u} +\mu B(\tilde{u})
\end{equation}
%%%%%
%%%%%
%%%%%
%%%%%
Let the function $\phi :{\rm I\!R}^{n_U}\rightarrow {\rm I\!R}^{n_U}$, depicted in Fig.~\ref{fig:phi} be  a family of MPCs, parametrised by $\mu$ (and $B$) generated by the nominal plant. The robustness of the system (\ref{model}) with additional unstructured uncertainty and controlled by (\ref{cycmin}) is considered using static and dynamic multipliers for IQCs for the barrier-based MPC.  For general $\mathcal{U}$, static multipliers are utilized; if $\mathcal{U}$ is time invariant (as is often the case) then the map $\phi$ is also time invariant so dynamic multipliers can be explored.
%%%%%-------------Preliminaries------------------------%%%%%%
\section{Background}\label{prelim}
%In this section, preliminary results required for this paper are presented.  Specifically, properties of the nonlinear functions are defined, the definition of integral quadratic constraints, the Zames-Falb multipliers and the stability theorem using the IQC's multipliers are presented. 
\subsection{Properties of Nonlinear Functions}
A multi-valued map $\phi$ is {\it sector-bounded} in the sense that there exists some $K \in {\rm I\!R}^{n\times n}$, $K > 0$ (or equivalently belongs to the sector $[0, K]$) such that
\begin{equation}\label{def_sector}
\phi(\theta)^T(K^{-1}\phi(\theta)-\theta)\leq 0 
\end{equation}
for all $\theta\in{\rm I\!R}^n$ and it is additionally {\it slope-restricted}, if there is $S\in {\rm I\!R}^{n\times n}$, $S> 0$ such that for all $\theta_x, \theta_y\in{\rm I\!R}^n$ and $\phi_x = \phi(\theta_x)$:
\begin{equation}\label{def_slope}
(\phi_y-\phi_x)^T(S^{-1}(\phi_y-\phi_x)-(\theta_y-\theta_x))\leq 0 
\end{equation}
Additionally, if $\phi(0) = 0$ then a slope-restricted nonlinearity is also sector bounded. Another property that is exploited for our main results is cyclic monotonocity. A  $n$-cyclic monotone increasing multi-valued map $\phi$ is defined as follows:
\begin{definition}
If $\phi$ is a $n$-cyclic monotone increasing map and $\phi_i = \phi(\theta_i)$ then $\forall n$
 \begin{equation}
%  \begin{split}
  \left\langle \theta_{0}-\theta_{1},\phi_{0} \right\rangle+\\
  \left\langle \theta_{1}-\theta_{2},\phi_{1} \right\rangle+ ...+\\
  \left\langle \theta_{n}-\theta_{0},\phi_{n} \right\rangle \geq 0
%  \end{split}
  \end{equation}
  \end{definition}
The $n$-cyclic monotone is an extension of the monotone property. Namely for $n=1$, inequality (11) turns into the monotone increasing property and the existence of a convex gradient function is summarized in~\cite{rockafellar1970convex}.
\begin{theorem}\label{rock}\cite{rockafellar1970convex}
Let $\phi$ be a multi-valued mapping from ${\rm I\!R}^{n}\rightarrow{\rm I\!R}^{n}$. In order for a closed proper convex function $P$ on ${\rm I\!R}^{n}$ such that $\phi(\theta)\subset\partial P$ for every $\theta$ to exist, it is necessary and sufficient that $\phi$ is cyclically monotone.
\end{theorem}
\subsection{Integral Quadratic Constraints}
IQC’s provide a way of conveniently representing associations between
nonlinear or possibly unknown processes~\cite{Megretski1997}. Two signals $w\in l_2^n$ and $\nu \in l_2^n$ are said to satisfy the IQC defined by a multiplier $\Pi(z)$, which is measurable, bounded and Hermitian, if
\begin{equation}
\int_{-\pi}^{+\pi}\begin{bmatrix}
\hat{w}(e^{j\omega})\\\hat{\nu}(e^{j\omega})
\end{bmatrix}^{*}
\Pi(e^{j\omega})
\begin{bmatrix}
\hat{w}(e^{j\omega})\\\hat{\nu}(e^{j\omega})
\end{bmatrix}d\omega\geq 0
\end{equation}
The classic stability theorem presented in~\cite{Megretski1997}, assumes that the interconnection between the system transfer function $G$ and the augmented nonlinearity $\Delta$ is well-posed. In addition the feedback interconnection between $G$ and $\Delta$ is stable if there exists $\epsilon>0$ such that
\begin{equation}\label{thm1}
\begin{bmatrix}
G(e^{j\omega})\\I
\end{bmatrix}^*\Pi(e^{j\omega})\begin{bmatrix}
G(e^{j\omega})\\I
\end{bmatrix}\leq -\epsilon I
\end{equation}
%
%\begin{theorem}\cite{Megretski1997}
%Let $G(z)\in \mathbf{RH}_{\infty}^{l\times m}$, and let $\Delta$ be a bounded causal operator. Assuming that:
%\begin{enumerate}
%\item for every $\tau\in$ [0,1], the interconnection of $G$ and $\tau\Delta$ is well-posed
%\item for every $\tau\in$ [0,1], the IQC defined by the multiplier $\Pi(z)$ is satisfied by $\tau\Delta$
%\item there exists $\epsilon>0$ such that\\
%\begin{equation}\label{thm1}
%\begin{bmatrix}
%G(e^{j\omega})\\I
%\end{bmatrix}^*\Pi(e^{j\omega})\begin{bmatrix}
%G(e^{j\omega})\\I
%\end{bmatrix}\leq -\epsilon I
%\end{equation}
%\end{enumerate}
%the feedback interconnection of $G$ and $\Delta$ is stable. 
%\end{theorem}
%
%In this case $\Delta$ represents the augmented nonlinearity, which consists of all the considered nonlinearity as in~\cite{Jonsson2000}
\subsection{Zames-Falb Multipliers}
%$Zames-Falb$ ({\it ZF}) are  special class of multipliers:
\begin{definition}\label{ZFMIMO}
The class of discrete-time rational Zames-Flab multipliers $\mathcal{M}$ contains all MIMO rational transfer functions $M_{ZF}\in \mathbf{RL}_{\infty}^{n\times n}$ such that $M_{ZF}(z)= H_s-H_{ZF}(z)$, where $||H_{ZF}||_1<   H_s$ are symmetric doubly hyper-dominant~\cite{willems1971analysis}:
\begin{equation}\label{full1}
\begin{split}
&H_{s_{ii}}\geq \sum_{j,j\neq i}|H_{s_{ij}}|+\sum_{j}||H_{ZF_{ij}}||_1
\end{split}
\end{equation}
with entries  $H_{ZF}\in l_1$.
Additionally, the subclass $\mathcal{M}_+\subset\mathcal{M}$ requires the following:
\begin{equation}
H_{s_{ij}}\leq0,H_{{ZF}_{ij}}\geq 0
\end{equation}
\end{definition}

%%%%%-------------End Preliminaries--------------------%%%%%%

\section{Properties for Barrier MPC}~\label{properties}
In this section the properties related to the  barrier MPC are explored to further derive the IQCs.  
%This formulation is proven to be advantageous for the improvement of input-output stability analysis, where both hard and relaxed satisfaction of the constraints can be used for the analysis. It is well known that when the barrier function is included in the control scheme, the solution changes according to the value of the parameter $\mu$, and the solution approximates the constrained solution when $\mu\rightarrow 0$; however this may produce numerical issues~\cite{Bazaraa2013}. The controller's input-output map satisfies IQC irrespective of $\mu$. This shows that same analysis can be done for all the positive values of $\mu$. 

The next two lemmas are required to show that $\phi: \theta_k \mapsto U_k$ (\ref{cycmin}) is slope-restricted, sector-bounded and cyclic monotone.

\begin{lemma}~\label{lmI}
\begin{enumerate}
    \item If $B \in \mathcal{B}$, then $\nabla B$ is monotone increasing and there exists $m\geq 0$ such that $\nabla^2B\geq m~I$  and $U^T\nabla B- m U^TU\geq 0$. \item If $B$ is also strongly convex then we can find $m>0$.
\end{enumerate}
\end{lemma}
\begin{IEEEproof}
%If $B\in\mathcal{B}$ is convex. Therefore, $\exists m\geq 0$ such that $\nabla^2 B\geq m I$. Then the function $B_{ex} = \nabla B-m U$ has a positive semi-definite gradient and $B_{ex}(0)=0$. Hence, $U^T\nabla B - m U^TU\geq 0$. {\color{red} If $B$ is strictly convex then $m>0$.}
\begin{enumerate}
    \item This is trivial with $m=0$ through convexity.
    \item If $B$ is strongly convex then we can find $m>0$ such that $\nabla ^2 B-mI\geq 0$. Define $\bar{B}=B-\frac{1}{2}mU^TU$. Then $\bar{B}$ is convex and the result follows since $U^T(\nabla B-mU) = U^T\nabla \bar{B}\geq 0$.
\end{enumerate}
\end{IEEEproof}

\begin{lemma}\label{remlmi}
If $\mathcal{U}$ is a set of bounded linear inequalities, $B \in \mathcal{B}$ and the constraints are either hard or $B$ is relaxed by a quadratic function $\beta_i$~\cite{Feller2015}, then the recentered barrier is strongly convex with $m>0$.
\end{lemma}
\begin{IEEEproof}
For linear and bounded constraints, there is a finite positive number $\delta_e$ such that $b_i-F_i(U)=b_i-L_iU\leq\delta_e$ . For a given $U\in\mathcal{U}$, the following holds:
\begin{equation}
\nabla^2{B}_i = \dfrac{L^T_iL_i}{(b_i-L_iU)^2}\geq  \dfrac{L^T_iL_i}{\delta_e^2}\geq 0
\end{equation}
It is trivial to show that there exits a $B_j$ such that $\nabla^2{B}_i+\nabla^2{B}_j>0$. 
%For hard constraints, $\nabla^2{B}_i\rightarrow +\infty$ when $U$ is approaching the boundary. In the case of relaxed constraints, for every $U$ that violates the $i^{th}$ constraint we have 
%\begin{equation}
%\nabla^2{B}_i = \nabla^2\beta_i = C_i>0
%\end{equation}
%whith $C_i$ being constant, as $\beta_i$ is quadratic.
\end{IEEEproof}
%For the rest of the paper the matrix $\tilde{H}$ as:
%\begin{equation}\label{Hhat}
%    \tilde{H} = H + \mu m I
%\end{equation}
%%%%
%%%%
%
%\begin{lemma}\label{mI}
%The relaxed gradient recentered barrier function~(\ref{eq1}) is strictly convex and there exists a parameter $m\geq 0$ such that $\nabla^2_UB\geq mI$, where $m$ is the least minimum eigenvalue. In addition, if $\mathcal{U}$ is linear and bounded then $B$ is strongly convex with $m>0$. 
%\end{lemma}  
\begin{theorem}\label{sector_nonlinear}
The nonlinearity $ \phi : {\rm I\!R}^{n_{U}} \rightarrow {\rm I\!R}^{n_U}$ (\ref{cycmin}) belongs to the sector [0,${\tilde{H}}^{-1}$], where $\tilde{H} = H + \mu m I$,  with $m\geq 0$ from lemma~\ref{lmI} and \ref{remlmi}, ${B}\in\mathcal{B}$ and $\mathcal{U} $is a convex set, if $\phi(0) = 0$.
\end{theorem}

\begin{IEEEproof}
%This is direct sequence of Eq.~\ref{def_sector} $\&$ \ref{def_slope} and Theorem~\ref{slope_proof}.
Using the KKT conditions  of (\ref{cycmin}) we have:
    \begin{equation}\label{sector}
        H\phi-\theta+\mu~\nabla B(\phi)=0\\
    \end{equation}
Since $\mathcal{U}$ is convex, multiplying (\ref{sector}) by $U^T$, using lemma~\ref{lmI} we get:
    \begin{equation}\label{sector2}
        \phi^T(H + \mu m I)\phi-\phi^T\theta\leq 0\\
    \end{equation}\end{IEEEproof}
\begin{theorem}~\label{slope_proof}
The nonlinearity $ \phi : {\rm I\!R}^{n_{U}} \rightarrow {\rm I\!R}^{n_U}$ (\ref{cycmin}) is additionally slope-restricted on [0,$\tilde{H}^{-1}$] with $m\geq 0$, ${B}\in\mathcal{B}$ and $\mathcal{U} $is a time invariant convex set.
\end{theorem}
\begin{IEEEproof}
Using the KKT conditions of (\ref{cycmin}) we have the following for $\phi_x = \phi(\theta_x)$ and $\phi_y = \phi(\theta_y)$
\begin{subequations}
    \begin{equation}\label{KKTm1}
        H\phi_x-\theta_x+\mu~\nabla B(\phi_x)=0\\
    \end{equation}
    \begin{equation}\label{KKTm2}
        H\phi_y-\theta_y+\mu~\nabla B(\phi_y)=0
    \end{equation}
\end{subequations}
Subtract (\ref{KKTm2}) from (\ref{KKTm1}) and multiply by $(\phi_y - \phi_x)^{T}$ to get:
\begin{equation}
\begin{split}
&(\phi_x-\phi_y)^{T} (H~(\phi_x-\phi_y)-(\theta_x-\theta_y))+\\
&+(\phi_x-\phi_y)^{T}(\mu~\nabla B(\phi_x)-\mu~\nabla B(\phi_y))=0\\
\end{split}
\end{equation}
%Since $\mathcal{U}$ is convex, then $\nabla{B}$ is monotone increasing. After 
Applying Lemma~\ref{lmI}\&\ref{remlmi}:
\begin{equation}\label{them1}
\begin{split}
&(\phi_x-\phi_y)^{T} (H ~(\phi_x-\phi_y)-(\theta_x-\theta_y))=\\
&-(\phi_x-\phi_y)^{T}(\mu~\nabla B(\phi_x)-\mu~\nabla B(\phi_y))\leq \mu m ||\phi_x-\phi_y||^2
%^T(\phi_x-\phi_y)\\
\end{split}
\end{equation}
Then 
\begin{equation}\label{them1a}
\begin{split}
&(\phi_x-\phi_y)^{T} ((H+\mu m I) ~(\phi_x-\phi_y)-(\theta_x-\theta_y))\leq 0\\
\end{split}
\end{equation}
Therefore, $\phi$ is slope-restricted on $[0,\tilde{H}^{-1}]$.
\end{IEEEproof}
%%%
%%%

\begin{figure}
\centering
\includegraphics[scale=0.1]{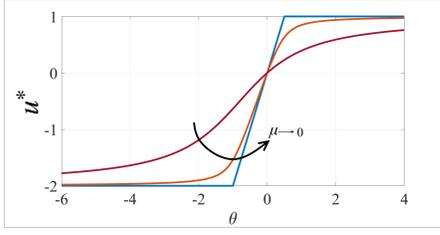}
\caption{Simple example to show that if a nonlinearity is parametrized by $\mu$ then its slope varies with $\mu$}
\label{fig:clas}
\end{figure}

%-----------------------%
It should be mentioned that for the sector bounded result there is no requirement for $\mathcal{U}$ being time-invariant. 
%--------------------------------%
%{\color{magenta} Don't use ``basically''. Make the next two paragraphs a formal Remark. And separate the ideas: (1) As you increase both $\mu$ and $m$ you widen the ? region. (2) for a simple system m corresponds to the slope. Example.... (3) Often m can be found analytically, as in the example, but sometimes not....}
\begin{remark}
This result shows that the inclusion of a barrier can change the maximum slope of the input-output map of the controller. This will widen the  stability region of the closed-loop system. It should be noted that $m$ depends only on the set of constraints and not on the design parameter $\mu$. In the numerical examples below, it will be shown that such a formulation can reduce the conservatism significantly in comparison to~\cite{Heath2006}.
\end{remark}
A simple example is utilized to illustrate the effect of the barrier-based MPC on the maximum slope. The nonlinearity is given by $U = (\arg\min_u 0.25u^2-\theta u+\mu (-ln(1-u)-ln(2+u)-0.5u)$. Fig.~\ref{fig:clas} depicts the solution for different values of $\mu$. In this case the value of $m$ (and the maximum slope) can be computed analytically (See Appendix), namely $m = 0.889$ and $Slope_{max} = (0.5+8/9\mu)^{-1}$.
%%%
%%%
Fig.~\ref{fig:clas} shows that the maximum slope decreases as $\mu$ increases. If $m$ is not computed (and is e.g. assumed to be $m=0$) then the maximum slope will be overestimated compared to the actual value of the slope, hence increasing the conservatism of the stability analysis. 
\begin{remark}
The parameter $m$ cannot always be calculated analytically. The search for $m$ may not be trivial and a signomial programming problem \cite{Boyd2007} is formulated such that the minimum eigenvalue of $\nabla^2 B$ is computed through deterministic methods. In $Appendix$ A it is shown that for the case of box constraints, $m$ can be calculated analytically. For general stage constraints the problem can be decomposed to smaller problems, solved using deterministic global optimization.
\end{remark}
%------------------------%%
The barrier MPC is cyclic monotone when the convex set $\mathcal{U}$ is time-invariant. 
\begin{theorem}\label{cycl_proof}
 When ${B}\in\mathcal{B}$ and  in addition $\mathcal{U} $is a time invariant convex set, then the nonlinearity $ \phi : {\rm I\!R}^{n_{U}} \rightarrow {\rm I\!R}^{n_U}$ (\ref{cycmin}) is $n-$cyclic monotone, 
\end{theorem}
\begin{IEEEproof}
For the barrier MPC the following is true
%  \begin{equation}
%  \begin{split}
%  &\phi_{0}^{T}(\theta_{0}-\theta_{1})+\phi_{1}^{T}(\theta_{1}-\theta_{2})+...+\phi_{n}^{T}(\theta_{n}-\theta_{0})=\\
%  &  \phi_{0}^{T}({H}(\phi_{0}-\phi_{1})+\mu (\nabla B(\phi_{0})-\nabla B(\phi_{1})))+\\
%  &  +\phi_{1}^{T}({H}(\phi_{1}-\phi_{2})+\mu (\nabla B(\phi_{1})-\nabla B(\phi_{2})))+\\
%  &  +...+\phi_{n}^{T}({H}(\phi_{n}-\phi_{0})+\mu (\nabla B(\phi_{n})-\nabla B(\phi_{0})))=\\
%  & \left[  \begin{matrix}
%    {\phi_{0}}&{\phi_{1}}& \dots&{\phi_{n}}\end{matrix}\right]
%A_c
%    \left[  \begin{matrix}
%    \phi_{0}&\phi_{1}& \dots&\phi_{n}\end{matrix}\right] ^{T}+\\
%   & +\mu(\nabla B(\phi_{0})^{T}(\phi_{0}-\phi_{n})+\nabla B(\phi_{n})^{T}(\phi_{n}-\phi_{n-1})+\\
%  &...+\nabla B(\phi_{1})^{T}(\phi_{1}-\phi_{0}))
%  \end{split}  
%  \end{equation}
     \begin{equation}
  \begin{split}
  &\sum_{k=0}^{n}\phi_{k}^{T}(\theta_{k}-\theta_{k+1})=\\
  & \sum_{k=0}^{n} \phi_{k}^{T}({H}(\phi_{k}-\phi_{k+1})+\mu (\nabla B(\phi_{k})-\nabla B(\phi_{k+1})))=\\
  & \left[  \begin{matrix}
    {\phi_{0}}&{\phi_{1}}& \dots&{\phi_{n}}\end{matrix}\right]
A_c
    \left[  \begin{matrix}
    \phi_{0}&\phi_{1}& \dots&\phi_{n}\end{matrix}\right] ^{T}+\\
   & +\mu\sum_{k=0}^{n}\nabla B(\phi_{k+1})^{T}(\phi_{k+1}-\phi_{k})
  \end{split}  
  \end{equation}
  %$$    A = \begin{bmatrix}
  %  H & -\dfrac{H}{2} & \dots & & &-\dfrac{H}{2}\\
  %  -\dfrac{H}{2} & H & -\dfrac{H}{2} & \dots& & 0\\
  %  0 & -\dfrac{H}{2} & H & -\dfrac{H}{2} & \dots & 0\\
  %  \vdots  & & & & & \vdots\\
  %  -\dfrac{H}{2} & 0 & \dots & & 0 & H
  %  \end{bmatrix}$$
  where $\phi_{n+1} = \phi_0$, $\theta_{n+1}=\theta_0$ and
  $$ A_c = \frac{1}{2}\begin{bmatrix}
        2H & - H & 0  & \cdots  &  & 0 & -H\\
        -H & 2H & -H &  &  & & 0\\
        0 & -H & 2H &    & & & \\
        \vdots & &    & \ddots &  & &  \vdots\\
         & &  &  &  2H & -H & 0\\
        0 & & & & -H & 2H & -H\\
        -H & 0 &  & \cdots   &  0 & -H & 2H
  \end{bmatrix}
  $$
  
 The matrix ${A}_c$ is always symmetric diagonally dominant with positive diagonal elements, thus it is positive semi-definite according to $\textit{Gershgorin circle theorem}$ \cite{varga}, so:
  \begin{equation}
  \left[  \begin{matrix}
    {\phi_{0}}&{\phi_{1}}& \dots&{\phi_{n}}\end{matrix}\right]
A_c
    \left[  \begin{matrix}
    \phi_{0}&\phi_{1}& \dots&\phi_{n}\end{matrix}\right] ^{T}\geq 0
  \end{equation}
Since $\textit{B}$ is convex its gradient is cyclic monotone: 
 \begin{equation}\label{lol}
 \begin{split}
 &\nabla B(\phi_{0})^{T}(\phi_{0}-\phi_{n})+\nabla B(\phi_{n})^{T}(\phi_{n}-\phi_{n-1})+...\\
 &+\nabla B^{T}(\phi_{1})(\phi_{1}-\phi_{0})\geq 0
 \end{split}
 \end{equation}
Thus the non-linearity is cyclic monotone.
\end{IEEEproof}

%%%
%%%
%%%-----------------End of Properties-----------------%%%%%
%%%-----------------Section for the IQCs--------------%%%%%
\section{Multipliers for Barrier MPC}\label{multipl}
IQCs for the barrier MPC will be derived in this section using the results from Section~\ref{properties}.
\subsection{Static Multipliers}
%Static multipliers for IQC can be found for a general convex set $\mathcal{U}$. 
\begin{corollary}~\label{thcircle}
For $\mathcal{U}$ being a convex set and  $B\in \mathcal{B}$, the nonlinearity $ \phi : {\rm I\!R}^{n_{U}} \rightarrow {\rm I\!R}^{n_U}$ (\ref{cycmin}), $\forall \theta \in l_2^{N_U}$admits IQC with the following mulitiplier  
\begin{equation}
%\int_{-\pi}^{+\pi}\begin{bmatrix}
%\hat{\theta}(e^{j\omega})\\\hat{\phi}(\theta)(e^{j\omega})
%\end{bmatrix}^{T}
\begin{bmatrix}
{0} & {I}\\
{I} & {-2\tilde{H}}
\end{bmatrix}
%\begin{bmatrix}
%\hat{\theta}(e^{j\omega})\\\hat{\phi}(\theta)(e^{j\omega})
%\end{bmatrix}d\omega\geq 0
\end{equation}
\end{corollary}
%with $m\geq 0$.
\begin{IEEEproof}
Immediate from Theorem~\ref{sector_nonlinear}. 
\end{IEEEproof}
%------------------------------%
\begin{remark}
 Theorem~\ref{sector_nonlinear} introduces an IQC for the barrier MPC where the constraints are generally convex constraints (as long as the optimization problem is always feasible). This case will be referred to as {\it general}. Nevertheless, by tightening the class of constraints, less conservative results can be found.
\end{remark}
\subsection{Dynamic Multipliers}
Time-invariant constraints can be used next to derive dynamic multipliers for less conservative stability analysis.
%%%%%%%%%%%%%%%%%%%%%%%%%%%%%%%%%%%%%%%%%%%%%%%%%%%%%%%%%%%%%%%%%
We prove the existence of ZF multipliers for the case of time-invariant convex constraints in Lemma \ref{cyccov}, and of less conservative multipliers for the case of box and staged constraints. %To avoid confusion in the annotation, the results will be given in terms of $\tilde{H}$ (\ref{Hhat}).
%%%-----------------------------------%%
%%%------------------------------------%%
%In order to provide the proof of existence of ZF multipliers the next lemma follow. 
According to~\cite{{Heath2007a},Safonov2000}, the existence of ZF multipliers additionally requires  the line integral $\int_A^B\phi(x)^Tdx$ to be independent of the path. This property is equivalent to the condition that $\phi$ is the gradient of some convex potential function. When, however,  the nonlinearity is not explicitly given, the above properties are difficult to be found. Here we propose the use of the property of cyclic monotonicity in order to prove the existence of the potential convex function.  
%KT: Check here
From Theorem~\ref{rock}, the conditions in ~\cite{Heath2007a,Safonov2000} can be substituted  by the condition of $\phi$ being cyclically monotone:
%%%%%%%%%%%%%%%%%%%%%%%%%%%%%%%%%%%%%%%%555
\begin{lemma}\label{cyccov}
Let $\phi:{\rm I\!R}^{n}\rightarrow {\rm I\!R}^{n}$  be  bounded  and  cyclic monotone increasing. Then for any $\theta\in l^n_2$ we have
\begin{equation}
\sum_{t={-\infty}}^{\infty} \theta(t+\tau)^{T}\phi (\theta(t))\leq \sum_{t={-\infty}}^{\infty} \theta(t)^{T}\phi (\theta(t))
\end{equation}
and if $\phi$ is odd then
\begin{equation}\label{odd}
\left|\sum_{t={-\infty}}^{\infty} \theta(t+\tau)^{T}\phi (\theta(t))\right|\leq \sum_{t={-\infty}}^{\infty} \theta(t)^{T}\phi (\theta(t))
\end{equation}
\end{lemma}
\begin{IEEEproof}
From Theorem~\ref{rock}, it is necessary and sufficient that the mapping is cyclically monotone, for a closed proper convex function $\textit{P}$ on ${\rm I\!R}^{n}$ to exist, such that $\phi \subset\partial P$. Now the results from~\cite{Heath2007a} can be employed in order to complete the proof.
\end{IEEEproof}
%%%%%%%
%%%%%%%
%%%%%%

%%%%%%
%%%%%%
%%%%%%
Next, the theorem for the existence of ZF mutipliers is given:
\begin{theorem}\label{ZF}
Let the nonlinearity $\phi$:$l_2^{n_{U}}\rightarrow l_2^{n_{U}}$  be bounded, $n$-cyclic monotone increasing and slope-restricted, with slope $\tilde{H}$. Let the SISO multiplier be $M\in \mathcal{M}_+$ (or $M\in \mathcal{M}$ and $\phi$ is additionally odd), then $M_{ZF} = M I$, and  $\forall \theta\in l_2^{n_U}$, $\phi$ admits IQCs with the following multipliers: 

%\begin{equation}\label{Mmon}
%\Pi(z)=
%\begin{bmatrix}
%{0} & {M_{ZF}(z)}^*\\
%{M_{ZF}(z)} & {0}
%\end{bmatrix}
%\end{equation}

\begin{equation}\label{Mslope}
\Pi(z) = 
\begin{bmatrix}
{0} & {M_{ZF}^*(z)}\\
{M_{ZF}(z)} & {-\tilde{H}M_{ZF}(z)-M_{ZF}^*(z)\tilde{H}}
\end{bmatrix}
\end{equation}
\end{theorem}
\begin{IEEEproof}
Use Lemma~\ref{cyccov} and~\cite{Heath2007a}. 
\end{IEEEproof}

\begin{remark}\label{ZFSISO}
If there is no special structure, e.g. repeated nonlinearities, then SISO multipliers should be used in the form of $M_{ZF} = M I$ with $M\in\mathcal{M}_{SISO}$. $\mathcal{M}_{SISO}~(or  \mathcal{M}_{{SISO}_+})$ contains all SISO rational transfer functions $M_{ZF}\in \mathbf{RL}_{\infty}$ that maintain the properties of Definition~\ref{ZFMIMO}.
\end{remark}
\begin{corollary}
The nonlinearity $ \phi : {\rm I\!R}^{n_{U}} \rightarrow {\rm I\!R}^{n_U}$ (\ref{cycmin}) admits IQC with multiplier $\Pi(z)$ (equation (\ref{Mslope})), $\forall \theta \in l_2^{N_U}$ with $\mathcal{U}$ a time-invariant convex set and  $B\in \mathcal{B}$.
\end{corollary}
\begin{IEEEproof}
\\
    (a)  The proof is almost identical with the one from \cite{Heath2007a}.
    %for the bound
    (b) It follows immediately from Theorem~\ref{cycl_proof} that $\phi$ is cyclic monotone.
    (c) It follows immediately from Theorem~\ref{slope_proof} that $\phi$ is slope-restricted on $[0, \tilde{H}^{-1}]$.
\end{IEEEproof}
The above results can be exploited to provide ZF multipliers when (possibly relaxed) barrier MPC is utilized.
\subsection{Multipliers for Box/Staged Constraints}
The conservatism can be dropped even further when a special structure of constraints (linear set) is used and more general multipliers than 
SISO ZF can be utilized. The following analysis generalizes the results from~\cite{Heath2010}, extending them to the case of (possibly relaxed) recentered Barrier MPC. The key idea is to represent the nonlinear function $\phi$ as an equivalent feedback structure. This structure is then modified to a nonlinear program $\psi$ together with a linear feedback term. The {\it new} nonlinear program $\psi$ can be separated into several smaller parallel nonlinear programs $\nu_i$. Multipliers can then be associated with each $\nu_i$. In our results, we show that there is a class of MIMO ZF multipliers $M_{ZF}$ for a special structure of staged and box constraints.% such that, the controller admits IQCs with the following multipliers.
%\begin{equation}\label{ZFM1}
%\Pi(z) = \begin{bmatrix} {0} & {M_{ZF}(z)^*}\\
%{M_{ZF}(z)} & {-\tilde{H}(K-H_{ZF}(z))-(K^T-H_{ZF}%(z))^*\tilde{H}}\end{bmatrix}%
%\end{equation} 
%\begin{equation}
%\langle\phi(\theta),Kx'~\rangle\geq \langle\phi(x),H_{ZF}*x'\rangle
%\end{equation} 
%%%%%
%%%%%%
%\subsubsection{Equivalent feedback structure} 
%%%%%

%The controller $\phi$ in Eq.~\ref{cycmin} can be transformed to an equivalent feedback structure. This will help to show that $\phi$ can be written as parallel nonlinear convex programs with a feedback loop. This equivalence will provide the preferable result.
Let $\psi:{\rm I\!R}^{n_U}\rightarrow {\rm I\!R}^{n_U}$ be the following convex program:
\begin{equation}\label{trans}
U=\psi(\theta')=\arg\min_u \dfrac{1}{2} u^{T}u-u^{T}\theta'+\mu B(u)
\end{equation} 
with $B\in \mathcal{B}$
\begin{lemma}\label{equi}
If $\theta'$ is equal to $\theta+(I-\tilde{H})\phi(\theta)$, then $U=\phi(\theta)$ and $U=\psi(\theta')$ are equivalent.
\end{lemma}
%%%%%%%%%%
\begin{IEEEproof}
Substituting $\theta'=\theta+(I-\tilde{H})\phi(x)$ in the KKT conditions of (\ref{trans}), our result follows immediately.
\end{IEEEproof}

\begin{figure}
\centering
\includegraphics[scale=0.25]{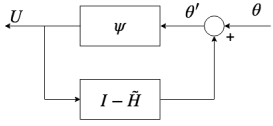}
\caption{Equivalent feedback structure}
\label{fig:psi}
\end{figure}

This equivalent feedback structure is depicted in Fig.~\ref{fig:psi} and a direct consequence of lemma~\ref{equi} is the following:
%%%%%%%%%%%%
\begin{equation}\label{transf}
\begin{bmatrix}\theta'\\U\end{bmatrix}=\begin{bmatrix}I&I-\tilde{H}\\{}&I\end{bmatrix}\begin{bmatrix}\theta'\\U\end{bmatrix}
\end{equation}
%%%%%%%%%%%%%%
%%%%%%%%%%%
%%%%%%%%%%%%%%%%%%
%%%%%%%%%%%%%%%%%%%
The rest of the analysis is based on the fact that $U=\psi(\theta')$ can be written as many parallel convex programs with $U=\sum_{i=0}^{N_{L-1}}u_i$. To do so, special structures of the constraints are considered such as limitations between adjacent actuators' movement (so-called staged constraints)~\cite{Heath2010} as well as box constraints. Both cases can be written as:
\begin{equation}\label{dec}
L=\begin{bmatrix}L_0^T & \dots & L_{N_L}^T\end{bmatrix}^T=diag(\tilde{L}_0, \dots,\tilde{L}_{N_L-1})
\end{equation}
%
%
%\begin{equation}\label{dec}
%L=\begin{bmatrix}L_0\\\vdots\\L_{N_L}\end{bmatrix}=\begin{bmatrix}\tilde{L}_0 &{}&{}\\
%{}&{\ddots}&{}\\{}&{}&\tilde{L}_{N_L-1}\end{bmatrix}
%\end{equation}
with $L_i$ having the following property:%%%%%%%Add the orthogonal stuff in the notation%%%%%
\begin{equation}\label{orthog}
L_iL_j^T=0
\end{equation}
for $i\neq j$. Hence, $U=\psi(\theta')$ can now be written as a set of parallel convex programs ($u_i$) for the case of staged and box constraints. 
%%%%%%LEMMA%%%%%
\begin{lemma}
The nonlinear convex program $U=\psi(\theta')$, given by (\ref{trans}),  can equivalently be transformed to $U=\sum_iu_i$, with $u_i$ being parallel convex programs:
\begin{equation} \label{eq11}
\begin{split}
&u_i=\arg\min_u\dfrac{1}{2}u^Tu-u^T\theta'+\mu \left( \sum_{j=1}^{N_0} \bar{B}_{ij}(u)+ln(b_{ij})-\dfrac{L_{ij}^T}{b_{ij}}u \right)\\
&s.t.~L_{i}^cu=0\\
&\bar{B}_{ij}(u)=\begin{cases}
-ln(b_{ij}-L_{ij}u)~for~ -L_{ij}U+b_{ij}\geq \delta\\
\beta_{ij}(u)~~~ elsewhere
\end{cases}
\end{split}
\end{equation}
\end{lemma}

\begin{IEEEproof}
See Appendix~B%\ref{a2}
\end{IEEEproof}

Now, let $\bar{L}_i$ be an orthonormal basis of the space spanned by the rows of $L_i$ and $\nu_i(p)$ be the convex program 
\begin{equation}
\nu_i(p) = \arg \min_q \dfrac{1}{2}q^Tq-q^Tp+\mu\left(\sum_{j=1}^{N_0} \bar{B}_{ij}\left(\bar{L}_{ij}^Tq\right)-\dfrac{\bar{L}_{ij}L_{ij}^T}{b_{ij}}q\right)
\end{equation}

A direct consequence is that each $\nu_i$ is bounded, n-cyclic monotone and slope restricted with slope $I$. Theorem~\ref{ZF} can, therefore, be applied:

\begin{lemma}\label{nu}
Let $M_{ZF_i}\in \mathcal{M}_+$ be a SISO rational strictly proper transfer function. Then  $\forall p\in l_2^{n_p}$, $\nu_i$ admits IQC with the following multiplier: 
\begin{equation}
{\Pi_{v}}_i(z)=\begin{bmatrix}
{0} & {M_{ZFi}^*}\\
{M_{ZFi}} & {- M_{ZFi}-M_{ZFi}^*}
\end{bmatrix}
\end{equation}
\end{lemma}

\begin{IEEEproof}
$\nu_i$ is bounded, $n$-cyclic monotone and slope restricted with slope $(I)$. Theorem~\ref{ZF} provides the result.
\end{IEEEproof}

The next lemma shows that each $u_i$ can be written as a function of $v_i$.
\begin{lemma}\label{eqmul}
Each $u_i$ can  equivalently be written as
\begin{equation}
u_i=\bar{L}_{i}^T\nu_i(\bar{L}_{i}\theta')
\end{equation}
\end{lemma}
\begin{IEEEproof}
See Appendix~C%\ref{a3}
\end{IEEEproof}

%The equivalence in lemma~\ref{eqmul} 
Consequently an IQC for the nonlinear system $\psi(\theta')$ with the following multiplier can be formulated:

\begin{equation}\label{psimul}
\Pi_\psi(z) = \sum_{i=0}^{N_L-1}\begin{bmatrix}\bar{L}_{i}&{}\\{}&\bar{L}_{i}\end{bmatrix}^T\Pi_{v_i}(z)\begin{bmatrix}\bar{L}_{i}&{}\\{}&\bar{L}_{i}\end{bmatrix}
\end{equation}

\begin{theorem}
 Let $\phi$:${l}_2^{n_{U}}\rightarrow l_2^{n_{U}}$ be bound, n-cyclic monotone increasing and slope-restricted, with slope $\tilde{H}$ under box or staged constraints. Let the multiplier $M_{ZF_i}\in \mathcal{M}_+$ be a SISO rational strictly proper transfer function,
$$M_{ZF}(z)=diag(M_{ZF_0}I, \dots, M_{ZF_{(N_L-1)}}I)$$
%\begin{bmatrix}M_{ZF_0}I&{}&{}\\{}&\ddots&{}\\{}&{}&M_{ZF_{(N_L-1)}}I\end{bmatrix} $$
%%%%%

then  $\forall \theta\in l_2^{n_U}$, $\phi$ admits IQC with the following multiplier:
 \begin{equation}\label{Multiplierzf}
\begin{split}
&\Pi(z)=\begin{bmatrix}
{I} & {I-\tilde{H}}\\
{0} & {I}
\end{bmatrix}^T\Pi_{\psi}(z)\begin{bmatrix}
{I} & {I-\tilde{H}}\\
{0} & {I}
\end{bmatrix}=\\
&\begin{bmatrix}
{0} & {M_{ZF}^*}\\
{M_{ZF}} & {-\tilde{H} M_{ZF}-M_{ZF}^*\tilde{H}}
\end{bmatrix}
\end{split}
\end{equation}

 \end{theorem}
 
 \begin{IEEEproof}
See Appendix~D%\ref{a4}
\end{IEEEproof}
 
 This theorem can be extended further for an even tighter class of box constraints where symmetric bounds are employed. That is, a wider class of multipliers can be employed as it can be proven that $\psi$ can be written as a linear transformation of repeated nonlinearities, hence full-block doubly hyper-dominant multipliers~\cite{willems1971analysis} can be utilized.
 
\iffalse 
 \begin{equation}
M_{ZF}(z)= H_s-H_{ZF}(z)
\end{equation}
%---------------------------%
with 
\begin{equation}\label{full}
\begin{split}
&H_{s_{ii}}\geq \sum_{j,j\neq i}|H_{s_{ij}}|+\sum_{j}||H_{ZF_{ij}}||_1\\
&H_{s_{ij}}\leq0
\end{split}
\end{equation}
\fi
%----------------------%
%In the next section, the convex search for the multipliers is presented.
%%%%%%%%%%%%%%%%%%%%%%%%%%%%%%%%%%%%%%%%%%%%%%%%%%%%%%%%%%%%%%%%
\section{Convex Search for Multipliers}\label{search}
In this section the convex search applied to the stability analysis is presented. %In order to conduct the appropriate stability analysis, linear matrix inequality (LMI) should be found in order to convert the test to the time domain using the KYP lemma~\cite{Rantzer2012}.
The results in this work allow the use of static multipliers for a wide class of constraints or dynamic multipliers for a tighter class of constraints. We revisit \cite{Fetzer2017b, Carrasco2018_search} and expand the results from \cite{Carrasco2018_search}, in order to incorporate a larger class of problems, where the slopes are given by a full-block matrix. %In the case of static multipliers it is straight forward as KYP lemma can directly be employed. However, for 
Dynamic multipliers may be non-causal and a factorization is required since they do not have a state-space representation. %, and the direct transformation to LMI is not direct.  
For the $i^{th}$ IQC its multiplier can be written as:

$\Pi_i(z) = \Psi^*(z)K_i \Psi(z)=\Psi^*(z)\begin{bmatrix}M_i^{11} & M_i^{12}\\M_i^{12T} & M_i^{22}\end{bmatrix}\Psi(z)$ 

For $N$ IQCs then we can then write: 
\begin{equation}
\Pi(z) = \Psi^*(z)K\Psi(z)
\end{equation}
\begin{equation}\label{PiIQC}
K=\begin{bmatrix}\begin{matrix}M_1^{11} & {} & {} \\{} & \ddots & {}\\ {} & {} & M_N^{11}\end{matrix} &\begin{matrix}M_1^{12} & {} & {} \\{} & \ddots & {}\\ {} & {} & M_N^{12}\end{matrix}\\\begin{matrix}M_1^{12T} & {} & {} \\{} & \ddots & {}\\ {} & {} & M_N^{12T}\end{matrix} & \begin{matrix}M_1^{22} & {} & {} \\{} & \ddots & {}\\ {} & {} & M_N^{22}\end{matrix}\end{bmatrix}
\end{equation}
In this work, finite impulse response (FIR) type multipliers (\ref{FIR}) are used. Their use can be justified using the phase equivalence argument~\cite{Carrasco2013}.

\begin{equation}\label{FIR}
M_{ZF} =\sum_{\substack{j=-N_{ZF_-}}}^{N_{ZF_+}}R_j(1-z^j)
\end{equation}
Using the largest $N_{ZF}=max(N_{ZF_+},N_{ZF_-})$ of all multipliers, $\Psi$ can be defined as: 

\begin{subequations}\label{psiz}
\begin{equation}
\Psi_{11} = \begin{bmatrix}I&(1-z^{-1})I&\dots&(1-z^{-N_{ZF}})I\end{bmatrix}^T
\end{equation}
\begin{equation}
\Psi = diag(\Psi_{11},\Psi_{11})%\begin{bmatrix}\Psi_{11}&\\
%&\Psi_{11}\end{bmatrix}
\end{equation}
\end{subequations}

The slope-restricted nonlinearities described in this work admit IQC multipliers as in (\ref{Multiplierzf}). As a result for the nonlinearity $\phi$, we have the following:
$M_\phi^{11} = 0$, $$M_{\phi}^{12} = \begin{bmatrix} R_0 & \dots & R_{N_{ZF}}\\\vdots \\ R_{-N_{ZF}}\end{bmatrix}$$ and 
$$M_\phi^{22} = \begin{bmatrix} R_0\tilde{H}+\tilde{H}R_0 & \dots & (R_{N_{ZF}}+R_{-N_{ZF}})\tilde{H}\\\vdots \\ \tilde{H}(R_{N_{ZF}}+R_{-N_{ZF}})\end{bmatrix}$$

%%%%%
\iffalse
The structure of $R_i$ depends on the method that is employed. In the case of C-ZF multipliers they are diagonal positive matrices.

Now, lets define $K= \sum_{i=0}^{N_L-1}\kappa _i L^T_iL_i=diag(\kappa _0 I,\dots,\kappa _{N_L-1}I)$ and
\begin{equation}
\begin{split}
H_{ZF} &= diag(h_0(z),\dots,h_{N_L-1}(z))\\
&=\sum_{\substack{i=-N_{ZF}\\i\neq0}}^{N_{ZF}}R_i(1-z^i)
\end{split}
\end{equation}
with all $R$ being diagonal matrices.
\fi
%%%%%

Therefore, the dynamic system $G_{\Psi}(z)$ has non-singular state-space representation and the LMI conditions can be constructed: 
\begin{equation}
G_{\Psi}(z)=\Psi(z)\begin{bmatrix}G(z)\\I\end{bmatrix}
\end{equation}
with:
\begin{equation}
G_{\Psi}\sim \left[
	\begin{array}{c|c}
	A_{\Psi} & B_{\Psi} \\
		\hline
    C_{\Psi} & D_{\Psi} 
	\end{array}
	\right]
\end{equation}
By the KYP lemma~\cite{Rantzer2012}, inequality~\ref{thm1} can be transformed into the following LMI optimization:
\begin{equation}\label{optim5}
\begin{split}
&\min_{\lambda,K}\lambda\\
&s.t.\\
&\begin{bmatrix}A_{\Psi}^TPA_{\Psi}-P & A_{\Psi}^TP B_{\Psi}\\B_{\Psi}^TP A_{\Psi} & B_{\Psi}^TP B_{\Psi}\end{bmatrix}+\\
&\begin{bmatrix}C_{\Psi} & D_{\Psi}\\0 & I \end{bmatrix}^TK\begin{bmatrix}C_{\Psi} & D_{\Psi}\\0 & I \end{bmatrix}\leq-\lambda I, \\
%&P>0\\
& and~additional~constraints~for~the~multipliers
\end{split}
\end{equation}
%---------------------------------------%
The additional constraints depend of the class of the multipliers (if $M\in\mathcal{M}_+ or \mathcal{M}$). For the case of static multipliers, $R_j$ for $j\neq 0$ can be set equal to zero. When diagonal multipliers are utilized (for example when asymmetric box or stage constraints are applied, $\it{C-ZF}$), then  $R_j\geq 0$ for $j\neq 0$ and $R_0>0$. The condition of doubly hyper-dominance can be similarly expressed.
%--------------------------------------%%---
\section{Robustness of Barrier MPC}\label{examples}
\begin{figure}
\centering
\includegraphics[scale=0.20]{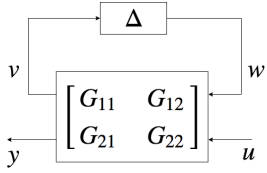}
\caption{Linear system under uncertainty}
\label{fig:linear}
\end{figure}

In this section the robustness of the barrier MPC is considered through an illustrative numerical example. 
If the transfer function of the open-loop LTI plant in (\ref{model}) is $G_{22}$, the system under the unstructured uncertainty ($\Delta$) is given by 
%
% KT: we use phi as the nonlinearity. Notation should be consistent. This needs to be rephrased or renoted
%
Fig.~\ref{fig:linear} with $\Delta: l^{n_\nu} \rightarrow l^{n_w}$. Then, the barrier MPC ($\phi$) can be included in the analysis as in \cite{Jonsson2000} with $\Delta = \begin{bmatrix}
\Delta_1 &\\&\phi
\end{bmatrix}$. The robustness of the MPC can be analyzed in terms of input-to-output stability given that the uncertainty admits an IQC. 
%---------------------------%
\subsection{Numerical Example}
The illustrative example consists of a nominal plant under a norm-bounded unstructured uncertainty $\Delta_1:l\rightarrow l$ which is bounded as $||\Delta_1||\leq b^2 $. The nominal dynamic system is given by the following equation:
\begin{equation}
\begin{split}
x_{k+1} &= \begin{bmatrix}{0.7} & {0.3}\\{0.8} & {0.01}\end{bmatrix}x_k+\begin{bmatrix}1\\0\end{bmatrix}u_k\\
y_k &= \begin{bmatrix}1&1.5\end{bmatrix}~x_k
\end{split}
\end{equation}
For this example the LTI plant has eigenvalues  0.9542, -0.2442 and zero equal to -1.19. As a result, the system is non-minimum phase.
 In addition to the nominal plant, the state observer is given by 
 %$J_u$ and $J_y$ represent 
\begin{equation}
\hat{x}(t) = J_u(z)u(t)+J_y(z)y(t)
\end{equation}
Here a steady-state Kalman filter is used with $J_u(z) = (zI-A+ALC)^{-1}B$ and $J_y = (zI-A+ALC)^{-1}AL$. For the numerical example the observer gain $L$ was calculated
via the discrete algebraic Riccati equation with weighting matrices
set equal to the identity matrix.

\begin{figure}[ht]
\centering
\includegraphics[scale=0.2]{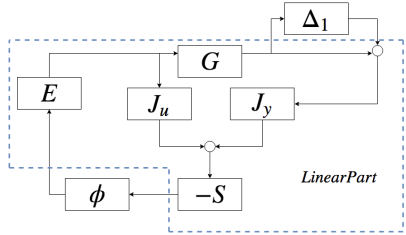}
\caption{Control Scheme}
\label{fig:clas1}
\end{figure}

%
%KT: which is the linear part in the Fig? It needs to be given here
The linear part of Fig.~\ref{fig:clas1} can be transformed into augmented linear system $M_s$ 
%
%KT: which is the "following for the linear part? 
%
\begin{equation}
M_s(z) = \begin{bmatrix}
\sqrt{b}~I& {}\\{}&-S
\end{bmatrix}
\begin{bmatrix}
{0} & {G}\\
{J_y} & {J_u+J_y G}
\end{bmatrix}\begin{bmatrix}
\sqrt{b}~I & {}\\{}&E
\end{bmatrix}
\end{equation}
with $I$ and $0$ the identity and zero matrix, respectively. Only the control action is applied and hence $E = \begin{bmatrix}
I & 0 & \cdots & 0
\end{bmatrix}$. Additionally, the scaled uncertainty is defined as $||\hat{\Delta}|| \leq 1$ and 
%
% KT: See previous comment about phi
%
$\phi$ the input-output map of the barrier MPC.
The control action is given by (\ref{cycmin}), where $\theta = -S~x$ and $B$ the gradient re-centered barrier function. 

\begin{figure}[ht]
\centering
\includegraphics[scale = 0.1]{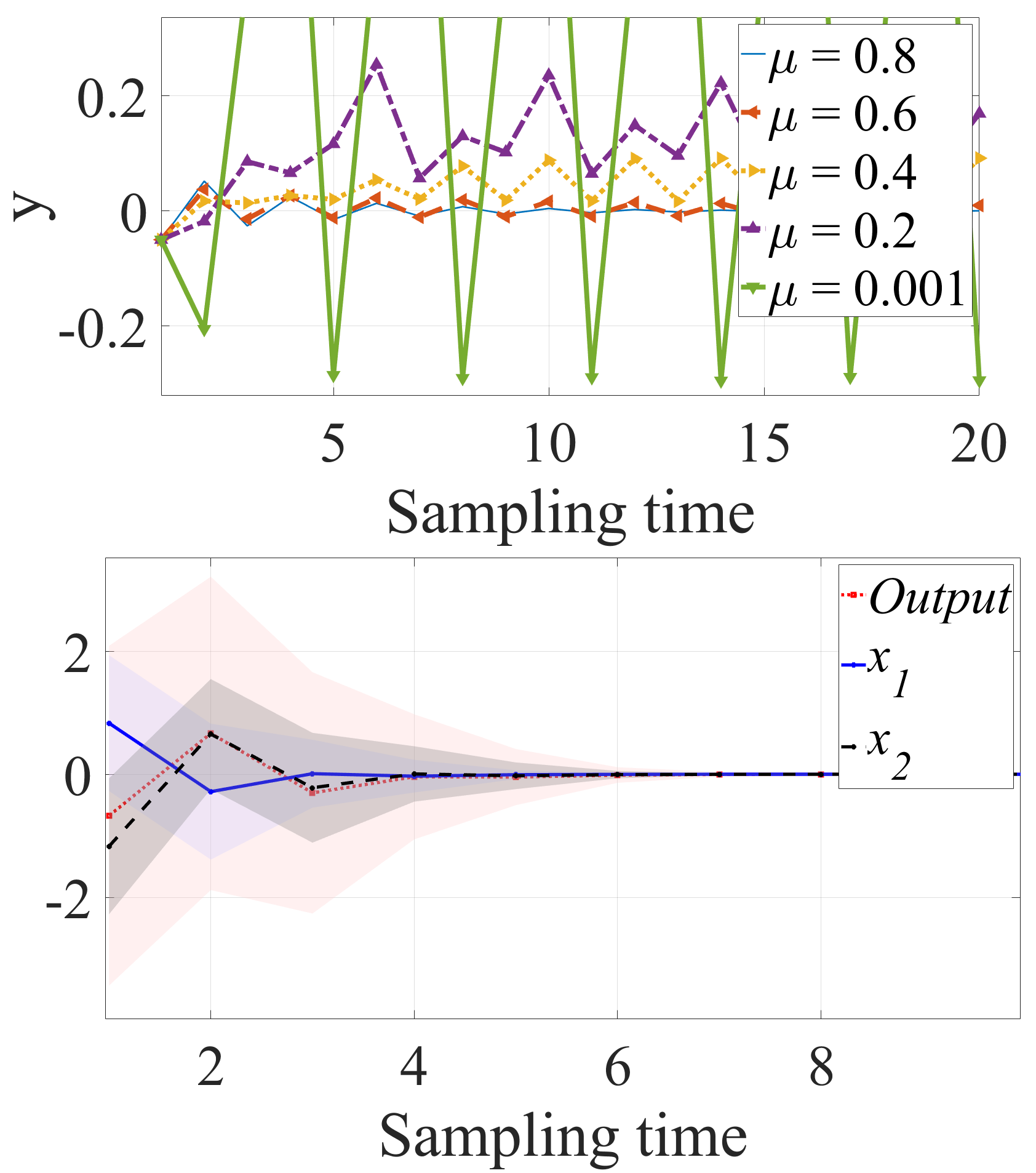}
\caption{Simulations of numerical example for (a) task 1 and (b) task 2}
\label{fig:ex2}
\end{figure}

As a result two IQCs can be written: one for the controller and one for the given unstructured uncertainty.  For $\hat{\Delta}$ we have $$\Pi_{\Delta} = \begin{bmatrix} I & 0\\0 & -I\end{bmatrix}$$ 
For the controller different multipliers are utilized depending on the case. 
All the algorithms presented in this work have been implemented for $\mu = 0.8$ and sufficiently large $N_{ZF}$. The constraints added to the manipulated variables are $-0.5\leq u_k\leq 1.0$. The control and prediction horizon are both set equal to 2 and $Q=I$. Two problems are investigated: {\it Task 1}. A positive gain $\kappa > 1$ is applied to the output of the dynamical system for $b = 0$ and the goal is to compute the maximum stable gain. {\it Task 2} is aimed at finding the smallest positive parameter $r$ of the objective function for $b=0.25$ and also the largest positive $b$ for $r = 0.001$ so that the system is guaranteed stable. The results for task 1 are shown in Table~\ref{results0} and the case of a nominal MPC is conducted for comparison purposes. From Table~\ref{results0}, the advantage of  barrier MPC compared  the nominal MPC becomes obvious. The maximum gain for the case of barrier MPC is 2.913 compared to 1.130 for the nominal MPC. Additionally, after trial and error, we found that for $\kappa = 3.4$, barrier MPC is destabilized, which is very close to the the computed value. Additionally, for $\kappa = 2.9$, some simulations have been conducted for various $\mu$. From Fig.~\ref{fig:ex2} (a) the advantage of the barrier MPC is clear, since the nominal MPC is unstable as expected since the maximum computed $\kappa$ is 1.130. 

\begin{table}[!ht]
\centering
\caption{Maximum $\kappa$ for stability}
\label{results0}
\begin{tabular}{lc}
\hline
{}       & $\begin{matrix} Nominal & Barrier\\ \end{matrix}$   \\    
\hline
General                  &  -~~~~~~ 1.091    \\
ZF ($N_{ZF} =10$)       & -~~~~~~ 1.091 \\
C-ZF       ($N_{ZF} =0$) & -~~~~~~ {2.539}  \\
C-ZF       ($N_{ZF} =1$)& 1.130  \textbf{2.913}  \\
C-ZF       ($N_{ZF} =10$)& 1.130  \textbf{2.913}\\
\hline
\end{tabular}
\end{table}
%%%%
For the next task, the results are depicted in Table~\ref{results1}. The analysis shows that barrier MPC is more robust than the nominal MPC, for all the different methods applied. Additionally, C-ZF seems to produce the least conservative results, predicting the system is stable for all possible $r$ even for $N_{ZF} = 1$.

\begin{table}[!ht]
\centering
\caption{Minimum $r$ for $b=0.25$ (a) and maximum $b$ for $r=0.1$ (b)}
\label{results1}
\begin{tabular}{lcc}
\hline
{} & ${\begin{matrix} (A)\\r~($for$~b = 0.25)\end{matrix}}$ & ${\begin{matrix} (B)\\b~($for$~r = 0.1)\end{matrix}}$\\% $b$  (for $r = 0.1$)}\\
\hline
{}       & $\begin{matrix} Nominal & Barrier\\ \end{matrix}$       &  $\begin{matrix} Nominal & Barrier\\ \end{matrix}$ \\
\hline
General                  &  3.994~ 1.150  &  $\begin{matrix} - &~~~~~ 0.0955 \end{matrix}$   \\
ZF ($N_{ZF} =1$)        & 3.568~ 0.724 & $\begin{matrix} - &~~~~~ 0.0986 \end{matrix}$  \\
ZF ($N_{ZF} =10$)       & 3.568~ 0.724 & $\begin{matrix} - &~~~~~ 0.0986 \end{matrix}$ \\
C-ZF       ($N_{ZF} =0$)                  & 1.963  \textbf{0.0001} &  $\begin{matrix} - &~~~~~ 0.3387 \end{matrix}$ \\
C-ZF       ($N_{ZF} =10$)& 0.098  \textbf{0.0001}& $\begin{matrix} 0.2510 & \textbf{0.5112}\end{matrix}$  \\
C-ZF       ($N_{ZF} =20$)& 0.098  \textbf{0.0001}&  $\begin{matrix} 0.2610 & \textbf{0.5112}\end{matrix}$ \\
\hline
\end{tabular}
\end{table}

Next, the design parameter is fixed at $r=0.1$ and parameter $b$ of the uncertainty changes. The results are depicted in Table~\ref{results1}~(B).
%KT: IT is not cleat what it is meant to say below
For this task, the robustness of the barrier MPC is again validated, where the most of the methods for nominal MPC failed to predict a stable system. However, all the stability tests performed for the barrier MPC predict a stable a region. 
%%%%%%%%%%%%%%%%%%%%%%%%%%%%%%%%%%
It should be mentioned that conservatism can be dropped even further by including more IQCs with respect to the uncertainty. 

To confirm that the closed loop system is stable a simulation is conducted for design parameter $r = 0.001$ and $b=0.25$ for several initial conditions. All of them produced stable results. These simulation results are depicted in Fig.~\ref{fig:ex2}(b). The lines represent the average behavior while the shaded areas correspond to the different initial values.

\section{Conclusion}~\label{conc}
In this work input-to-output stability results are developed the barrier MPC. The barrier can improve the robustness of the MPC due to the change in the slope. Additionally, general convex constraints can be employed. Tighter time-invariant convex constraints as well as staged constraints are considered. The tighter the constrained case the less conservative the analysis can become, through the use of dynamic multipliers. A convex search is presented in order to be able to apply the stability criteria using multipliers. The results of this paper can be further generalized for the case of PWA models extending our recent work~\cite{PETSAGKOURAKIS2018_pwa_barrier}.

\section*{Acknowledgment}
University of Manchester Presidential Doctoral Scholarship
Award to PP is gratefully acknowledged.
\bibliographystyle{IEEEtran}
\bibliography{IEEEabrv,IEEEexample}
\appendices
\section{}\label{a1}
The parameter $m$ that makes the slope of the nonlinearity tighter seems to be able to significantly reduce conservatism. The corresponding optimization problem is not convex and this may complicate the algorithm. Nevertheless, here we demonstrate that even though the problem is not convex it can have only one solution in the feasible region. The parameter $m$ can be calculated by minimizing the smallesty eigenvalue of the Hessian of the barrier inside the feasible region. This problem can be formulated as:
\begin{equation}\label{eigp}
\begin{split}
m = \min_{u,x} x^T\nabla^2_uB(u)x\\
s.t.~x^T x=1\\
 Lu\leq b
 \end{split}
\end{equation}
where $$\nabla^2_uB(u) = \sum_i \dfrac{1}{(b_i-L_i u)^2}L_i^TL_i$$
The KKT conditions of this problem can be written as the following set of equations:
\begin{subequations}\label{apkkt}
\begin{align}
 &\sum_i \dfrac{1}{(b_i-L_i u)^2}L_i^TL_ix-\lambda~x=0\\
 &1-x^T x = 0\\
 &\sum_i \dfrac{1}{(b_i-L_i u)^3}L_i^T x L_i^TL_ix-\sum_i\dfrac{\lambda^{in}_i}{2} L^T_i=0\label{apkktt3}\\
& \lambda^{in}_i (L^T_i u-b) = 0
\end{align}
\end{subequations}
This set of equations corresponds to the minimum of the smallest eigenvalue. The optimum objective function will be equivalent to the parameter $m$ used in our analysis. This optimization, however,  as is non-convex in the general case can be a bottleneck for the conservatism of the proposed analysis. Here a result regarding the box constraints is provided as well as a relaxation for the general case of bounded constraints. Box constraints can be seen as

\begin{equation}
L = \begin{bmatrix}
1 & 0 & \dots & 0\\-1 & 0 & \dots & 0\\
0 & 1 & \dots & 0\\0 & -1 & \dots & 0\\
& \vdots &\\ 
0 & 0 & \dots & 1\\0 & 0 & \dots & -1\end{bmatrix}\end{equation}

\begin{equation}
b = \begin{bmatrix}
\bar{b}_1\\
\underline{b}_1\\
\vdots\\
\end{bmatrix}\end{equation}

(\ref{apkktt3}) of the KKT conditions assuming the solution can never be in the bound of the polyhedral, can be written as:
\begin{equation}\label{polyhed}
 -\dfrac{1}{(\underline{b}_i +u_i)^3}\begin{bmatrix}0\\\vdots \\x_i^2\\\vdots\\0\end{bmatrix}+\dfrac{1}{(\bar{b}_i -u_i)^3}\begin{bmatrix}0\\\vdots \\x_i^2\\\vdots\\0\end{bmatrix}=0
\end{equation}
The solution of \ref{polyhed} is unique and independent of $x$. Precisely, the value of $m$ is calculated as: $\min[{\dfrac{8}{(\bar{b}_1+\underline{b}_1)^2}},\dots,{\dfrac{8}{(\bar{b}_i+\underline{b}_i)^2}},\dots,{\dfrac{8}{(\bar{b}_N+\underline{b}_N)^2}}]$. .

Nevertheless, the previous case accounts only box constraints, and a decomposition for the staged constraints is next considered. The Hessian can be decomposed using (\ref{dec}) as:
\begin{equation}\label{hes}
\nabla^2_uB(u) = \begin{bmatrix}
\sum_{i=1}^{n_u}\dfrac{\tilde{L}_0^T\tilde{L}_0}{(\tilde{b}_0-\tilde{L}_{0i}u_0)^2} & 0 & \cdots & 0 \\ 0 &
\sum_{i=1}^{n_u}\dfrac{\tilde{L}_1^T\tilde{L}_1}{(\tilde{b}_1-\tilde{L}_{1i}u_1)^2} & \cdots & 0\\ 0 & 0 & 0 & \ddots 
\end{bmatrix}
\end{equation}

Therefore, since (\ref{hes}) has block diagonal structure, the eigenvalues can be found separately for each block using (\ref{eigp}), and the smallest one can be selected as $m$
\section{}\label{a2}
The KKT conditions for $u_i$ with $i=0,...,N_L-1$ are
\begin{equation}\label{eq2}
\begin{split}
u_i-\theta'+\mu\left(\sum_{j=1}^{N_0}\nabla \bar{B}_{ij}-\dfrac{L_{ij}^T}{b_{ij}} \right)+L_i^{cT}z_i=0\\
L_i^cu_i=0
\end{split}
\end{equation}
with
\begin{equation}\label{z}
z_i=-L_i^c\theta' 
\end{equation}
for $i=0..N_L-1$. Summing \ref{eq2} over $i$ together with \ref{z} gives
\begin{equation}
\begin{split}
&u-(N_L-1)\theta'+\sum_{i=0}^{N_L-1}\mu\left(\sum_{j=1}^{N_0}\nabla\bar{B}_{ij}(u_i)-\dfrac{L_{ij}^T}{b_{ij}}  \right)=\\
&=\sum_{i=0}^{N_L-1}L_i^{cT}L_i^c\theta'=(N_L-2)\theta'
\end{split}
\end{equation}
Therefore
\begin{equation}
u-\theta'+\mu\nabla B(u)=0
\end{equation}
\section{}\label{a3}
$\nu_i$ is bounded, $n$-cyclic monotone and slope restricted with slope $(I)$. Corollary~\ref{ZF} gives the result.

\section{}\label{a4}

$\phi(\theta)$ can be expressed with respect to $\psi(\theta')$, using (\ref{transf}), as a result $$\Pi(z)=\begin{bmatrix}
{I} & {I-\tilde{H}}\\
{0} & {I}
\end{bmatrix}^T\Pi_{\psi}(z)\begin{bmatrix}
{I} & {I-\tilde{H}}\\
{0} & {I}
\end{bmatrix}$$
To complete the proof, we use the fact that $$\bar{L}_i^T\bar{L}_i= \begin{bmatrix}0&{}&{}&{}\\{}&{\ddots}&{}\\{}&{}&I&{}\\{}&{}&{}&{\ddots}&{}\\{}&{}&{}&{}&0\end{bmatrix}\Bigg\}i^{th}~row$$

\end{document}